\documentclass[aps,pra,floatfix,showpacs,preprint,superscriptaddress]{revtex4-2}
\usepackage{bm}
\usepackage{mathrsfs}
\usepackage{amsmath}
\usepackage{amssymb}
\usepackage{revsymb}
\usepackage{accents}
\usepackage{color}
\usepackage[force]{feynmp-auto}
\usepackage{hyperref}
\usepackage{cleveref}

\usepackage{graphicx}

\begin{document}
\title{First-order strong-field QED processes including the damping of particles states}
\author{T. \surname{Podszus}}
\author{A. \surname{Di Piazza}}
\email{dipiazza@mpi-hd.mpg.de}
\affiliation{Max Planck Institute for Nuclear Physics, Saupfercheckweg 1, D-69117 Heidelberg, Germany}
\date{\today}

\begin{abstract}
Volkov states are exact solutions of the Dirac equation in the presence of an arbitrary plane wave. Volkov states, as well as free photon states, are not stable in the presence of the background plane-wave field but ``decay'' as electrons/positrons can emit photons and photons can transform into electron-positron pairs. By using the solutions of the corresponding Schwinger-Dyson equations within the locally-constant field approximation, we compute the probabilities of nonlinear single Compton scattering and nonlinear Breit-Wheeler pair production by including the effects of the decay of electron, positron, and photon states. As a result, we find that the probabilities of these processes can be expressed as the integral over the light-cone time of the known probabilities valid for stable states per unit of light-cone time times a light-cone time-dependent exponential damping function for each interacting particle. The exponential function for an incoming (outgoing) either electron/positron or photon at each light-cone time corresponds to the total probability that either the electron/positron emits a photon via nonlinear Compton scattering or the photon transforms into an electron-positron pair via nonlinear Breit-Wheeler pair production until that light-cone time (from that light-cone time on). It is interesting that the exponential damping terms depend not only on the particles momentum but also on their spin (for electrons/positrons) and polarization (for photons). This additional dependence on the discrete quantum numbers prevents the application of the electron/positron spin and photon polarization sum-rules, which significantly simplify the computations in the perturbative regime.
  
\end{abstract}
 
\maketitle

\section{Introduction}

There is a growing interest in testing QED under the extreme conditions provided by intense laser fields 
\cite{Mitter_1975,Ritus_1985,Ehlotzky_2009,Reiss_2009,Di_Piazza_2012,Dunne_2014,Blackburn_2020}. The typical electromagnetic field scale of QED is determined by the so-called ``critical'' field of QED: $F_{cr}=m^2/|e|=1.3\times 10^{16}\;\text{V/cm}=4.4\times 10^{13}\;\text{G}$ (we employ units with $\epsilon_0=\hbar=c=1$ throughout, and $m$ and $e<0$ denote the electron mass and charge, respectively) \cite{Landau_b_4_1982,Fradkin_b_1991,Dittrich_b_1985}. In general, in the presence of electromagnetic fields of the order of $F_{cr}$ the vacuum becomes unstable under electron-positron pair production and the magnetic interaction energy associated with the intrinsic electron magnetic moment becomes of the order of the electron rest energy.

High-power optical lasers are becoming an important tool to test QED at critical field strengths, which correspond to laser intensities of the order of $10^{29}\;\text{W/cm$^2$}$. Due to the Lorentz-invariance of the theory, in fact, observable quantities like probabilities and rates of physical processes, depend on Lorentz-invariant parameters. For processes initiated by an electron/positron (a photon) with four-momentum $p^{\mu}=(\varepsilon,\bm{p})$ ($q^{\mu}=(\omega,\bm{q})$), with $\varepsilon=\sqrt{m^2+\bm{p}^2}$ ($\omega=|\bm{q}|$) in the presence of a background field with amplitude given by the electromagnetic field tensor $F_0^{\mu\nu}=(\bm{E}_0,\bm{B}_0)$ in the laboratory frame, the invariant parameter characterizing the strength of the field is the so-called quantum nonlinearity parameter $\chi_0=\sqrt{|(F_0^{\mu\nu}p_{\nu})^2|}/mF_{cr}$ ($\kappa_0=\sqrt{|(F_0^{\mu\nu}q_{\nu})^2|}/mF_{cr}$), with the metric tensor $\eta^{\mu\nu}=\text{diag}(+1,-1,-1,-1)$ \cite{Mitter_1975,Ritus_1985,Ehlotzky_2009,Reiss_2009,Di_Piazza_2012,Dunne_2014}. In the case of an incoming electron/positron, this parameter corresponds to the (electric) field strength in the rest frame of the particle. Thus, although available lasers have reached peak intensities $I_0$ of the order of $5.5\times 10^{22}\;\text{W/cm$^2$}$ \cite{Yoon_2019} and upcoming facilities aim at $I_0\sim 10^{23}\text{-}10^{24}\;\text{W/cm$^2$}$ \cite{APOLLON_10P,ELI,CoReLS,Bromage_2019,XCELS}, the availability of ultrarelativistic electron/positron beams allows for testing the theory effectively at the critical field scale \cite{Mitter_1975,Ritus_1985,Ehlotzky_2009,Reiss_2009,Di_Piazza_2012,Dunne_2014}. At the mentioned available intensities, in fact, an electron with an energy of the order of $1\;\text{GeV}$, already within the reach of present technology, would experience a field of the order of $F_{cr}$ in its rest frame.

First experiments in this ``strong-field'' regime of QED have been performed at SLAC in the late nineties \cite{Bula_1996,Burke_1997,Bamber_1999} and recently two experiments have been also carried out probing laser-electron interaction at values of the quantum nonlinearity parameter close to unity \cite{Cole_2018,Poder_2018}. Also, devoted experimental campaigns are already planned at DESY \cite{Abramowicz_2019} and at SLAC \cite{Meuren_2020} to test QED in the strong-field regime.

On the theory side the description of the interaction of high-intensity optical lasers, as those mentioned above, and electrons/positrons is complicated because the density of laser photons is so high that nonlinear effects in the laser electromagnetic field amplitude play a major role \cite{Mitter_1975,Ritus_1985,Ehlotzky_2009,Reiss_2009,Di_Piazza_2012,Dunne_2014}. These effects are controlled by the classical nonlinearity parameter $\xi_0=|e|E_0/m\omega_0$, where $\omega_0$ is the central angular frequency of the laser field. The parameter $\xi_0$ does not contain the Planck constant and classically controls the importance of relativistic effects in laser-electron/positron interaction. For optical lasers the parameter $\xi_0$ exceeds unity already at laser intensities of the order of $10^{18}\;\text{W/cm$^2$}$ and for $\xi_0\gtrsim 1$ the laser-electron/positron interaction has to be taken into account exactly in the calculations. This is achieved in QED within the so-called Furry picture \cite{Furry_1951}, where the electron-positron field is quantized in the presence of the background laser field \cite{Fradkin_b_1991,Landau_b_4_1982}. This in turn requires that the Dirac equation can be solved analytically in the presence of the background field, which has been carried out in Ref. \cite{Volkov_1935} in the case of a plane wave (see also Ref. \cite{Landau_b_4_1982}). The corresponding electron/positron states (and propagator) are known as Volkov states (Volkov propagator).

By employing the Volkov states, the basic processes corresponding to the emission by an electron/positron of a single photon (nonlinear Compton scattering) and to the transformation of a photon into an electron-positron pair (nonlinear Breit-Wheeler pair production) have been extensively investigated (see Refs. \cite{Goldman_1964,Nikishov_1964,Ritus_1985,Baier_b_1998,Ivanov_2004,Boca_2009,
Harvey_2009,Mackenroth_2010,Boca_2011,Mackenroth_2011,Seipt_2011,Seipt_2011b,
Dinu_2012,Krajewska_2012,Dinu_2013,Seipt_2013,
Krajewska_2014,Wistisen_2014,Harvey_2015,Seipt_2016,Seipt_2016b,Angioi_2016,
Harvey_2016b,Angioi_2018,Di_Piazza_2018,Alexandrov_2019,Di_Piazza_2019,Ilderton_2019_b} for nonlinear Compton scattering and Refs. \cite{Reiss_1962,Nikishov_1964,Narozhny_2000,Roshchupkin_2001,Reiss_2009,
Heinzl_2010b,Mueller_2011b,Titov_2012,Nousch_2012,Krajewska_2013b,Jansen_2013,
Augustin_2014,Meuren_2016,Di_Piazza_2019,King_2020} for nonlinear Breit-Wheeler pair production, as well as the reviews \cite{Mitter_1975,Ritus_1985,Ehlotzky_2009,Reiss_2009,Di_Piazza_2012}). 

Now, if one computes the total probabilities of nonlinear Compton scattering and nonlinear Breit-Wheeler pair production at the leading order in perturbation theory, one observes that for sufficiently long pulses they can exceed unity. The reason behind this apparent contradiction relies on the importance of higher-order processes. This is, for example, intuitively clear in the case of nonlinear Compton scattering as for sufficiently long pulses the probability that electrons/positrons emit a higher number of photons becomes sizable. In Ref. \cite{Glauber_1951}, Glauber has shown that in the classical limit of nonlinear Compton scattering, where recoil effects are negligible, the emission of an arbitrary number of photons by an electron is described by a Poisson distribution. Relying on the unitarity of the $S$-matrix of QED, this result has been obtained by imposing that the total probability that an electron either does not emit a photon or does emit an arbitrary number of photons is equal to unity. In Ref. \cite{Di_Piazza_2010} the same idea was applied in strong-field QED in the so-called moderately quantum regime where $\xi_0\gg 1$ and $\chi_0\lesssim 1$ such that nonlinear Breit-Wheeler pair production was negligible and the so-called locally-constant field approximation (LCFA) was employed \cite{Ritus_1985,Baier_b_1998,Di_Piazza_2012}. From a QED point of view, the prescription used in Refs. \cite{Glauber_1951,Di_Piazza_2010} was phenomenological and not based on first principles. Indeed, from the unitarity of the $S$-matrix it has to follow automatically that probabilities of physical processes never exceed unity. An alternative, consistent approach in this respect was then presented in Ref. \cite{Neitz_2013} but based on distribution functions and kinetic equations rather than on single-particle probabilities (the inclusion of the process of pair production was carried out in Ref. \cite{Neitz_2014} and we stress here that kinetic equations had already been used in strong-field QED to describe the formation and the evolution of QED cascades \cite{Elkina_2011,Nerush_2011}). The problem of radiation of several photons is closely related to the problem of radiation reaction in QED, which has also a classical counterpart \cite{Landau_b_2_1975,Jackson_b_1975,Rohrlich_b_2007}. Indeed, the inclusion of classical radiation-reaction effects in the computation of emission spectra has been investigated numerically in several works \cite{Hartemann_1996,Koga_2004,Di_Piazza_2009,Lehmann_2011,Harvey_2011,Bulanov_2011,Kumar_2013,
Capdessus_2014,Tamburini_2014}. Moreover, the availability of the exact solution of the underlying classical equation of motion including radiation reaction (the Landau-Lifshitz equation \cite{Landau_b_2_1975,Rohrlich_b_2007}) \cite{Di_Piazza_2008_a} has also allowed one to obtain analytical results on the classical emission spectra including radiation reaction \cite{Di_Piazza_2018_b,Heinzl_2021,Di_Piazza_2021}.

We have mentioned that the unitarity of the $S$-matrix guarantees that computed probabilities do not exceed unity. However, this implication holds for probabilities computed \emph{exactly} and the use of perturbation theory may and does lead to violations of unitarity. The contradictions are only apparent because the use of perturbation theory is allowed only in those regimes where the obtained probabilities are smaller than unity. In this respect, a refined probabilistic approach has been presented in Ref. \cite{Tamburini_2019} to show that exact radiation probabilities feature a time-dependent exponential suppression related to the fact that electron/positron Volkov states are not stable states due to the emission of photons. 

From the point of view of strong-field QED, in order to compute, for example, the exact probability of processes like nonlinear Compton scattering and nonlinear Breit-Wheeler pair production, one has to use the exact electron and photon states in a plane wave including radiative corrections as well as the exact expression of the vertex. In order to compute the exact electron/positron and photon states in a background electromagnetic field described by the four-vector potential $A^{\mu}(x)$, one has to solve the corresponding Schwinger-Dyson equations
\begin{align}
\label{SD_psi}
\{\gamma^{\mu}[i\partial_{\mu}-eA_{\mu}(x)]-m\}\Psi(x)&=\int d^4y\, M(x,y)\Psi(y),\\
\label{SD_a}
-\partial_{\mu}\partial^{\mu}\mathscr{A}^{\nu}(x)&=\int d^4y\, P^{\nu\lambda}(x,y)\mathscr{A}_{\lambda}(y),
\end{align}
where $\gamma^{\mu}$ are the Dirac matrices, $\Psi(x)$ is the electron-positron field, $\mathscr{A}^{\mu}(x)$ is the radiation field in the Lorenz gauge, and where $M(x,y)$ and $P^{\mu\nu}(x,y)$ are the exact mass and polarization operator in the external field \cite{Landau_b_4_1982}. The mass operator and the polarization operator correspond to the sum of all possible one-particle irreducible Feynman diagrams with two external electron/positron and photon lines. The contribution of the one-particle reducible diagrams is, instead, exactly taken into account by the Schwinger-Dyson equations themselves, which can be seen by writing the solutions of Eqs. (\ref{SD_psi})-(\ref{SD_a}) as a perturbative series in $M(x,y)$ and $P^{\mu\nu}(x,y)$, respectively (see also below). On the contrary, the exact vertex does not feature by definition one-particle reducible contributions. This is an important remark for what it follows because in the presence of a plane wave, probability amplitudes receive also contributions for momentum regions where electron and photon propagators describing internal lines are on-shell (we are referring here to amplitudes corresponding to Feynman diagrams which split into two diagrams by cutting the corresponding internal line). This is a consequence, ultimately, of the fact that, unlike in vacuum, single-vertex processes like nonlinear Compton scattering and nonlinear Breit-Wheeler pair production do occur in the presence of the plane wave. This aspect has been thoroughly investigated in the study of higher-order strong-field QED processes in a plane wave like the emission of two photons by an electron (nonlinear double Compton scattering) \cite{Loetstedt_2009,Seipt_2012,Mackenroth_2013,King_2015,Dinu_2019}, the emission by an electron of a photon, which then decays into an electron-positron pair (nonlinear trident pair production) \cite{Hu_2010,Ilderton_2011,King_2013,Dinu_2018,Mackenroth_2018,Dinu_2020,Torgrimsson_2020}, and the annihilation into two photons of an electron-positron pair \cite{Bragin_2020}. In these studies the contribution to the probabilities stemming from intermediate on-shell particles has been indicated as ``incoherent'' or ``two-step'' contribution, and it features a quadratic dependence on the laser pulse duration rather than linear as the remaining ``coherent'' or ``one-step'' contribution. Indeed, the quadratic dependence is easily understood as arising from the fact that the two single-vertex strong-field QED processes building the whole second-order process can occur independently and at any phase of the plane wave. For two-vertex processes primed by a single particle and for both the classical and the quantum nonlinearity parameters being of the order of unity, the one-step (two-step) contribution has been found to scale as $\alpha^2\Phi_L$ ($\alpha^2\Phi^2_L$), where $\alpha=e^2/4\pi\approx 1/137$ is the fine-structure constant and $\Phi_L$ is the total phase duration of the plane wave. Thus, for sufficiently long pulses ($\Phi_L\gg 1$) not-only the two-step contribution dominates over the one-step contribution but for $\Phi_L\gtrsim 1/\alpha\approx 137$ the probability of a two-step, second-order process would become comparable with that of a first-order process, a condition already identified in Ref. \cite{Di_Piazza_2010}. This circumstance is already reflected by the Poisson distribution of the number of photons emitted as found in Ref. \cite{Glauber_1951} and it also occurs in the case of radiative corrections. The structure of the Schwinger-Dyson equations (\ref{SD_psi})-(\ref{SD_a}) gives the possibility of taking into account these ``accumulation'' effects in the electron/positron and photon states exactly and self-consistently. In fact, the incoherent contribution to the one-particle reducible diagrams which can be cut into two diagrams at $n$ internal lines correspond to terms scaling with the $n$th power of the pulse duration and, as we have already mentioned, all these contributions are self-consistently ``resummed'' in the Schwinger-Dyson equations. 

\begin{figure}
\begin{center}
\includegraphics[width=0.6\columnwidth]{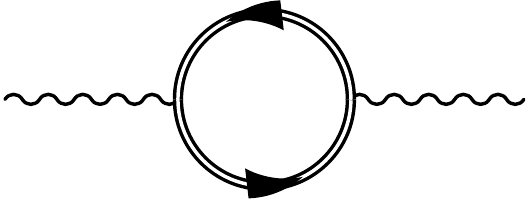}
\caption{The one-loop polarization operator in an intense plane wave. The double lines represent exact electron propagators in the plane wave (Volkov propagators) \cite{Landau_b_4_1982}.}
\label{FD_PO}
\end{center}
\end{figure}

\begin{figure}
\begin{center}
\includegraphics[width=0.4\columnwidth]{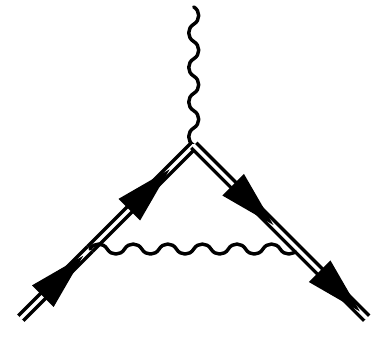}
\caption{The one-loop vertex correction in an intense plane wave. The double lines represent exact electron states and propagator in the plane wave (Volkov states and propagator, respectively) \cite{Landau_b_4_1982}.}
\label{FD_VC}
\end{center}
\end{figure}

It is important to stress that the mentioned accumulation effects do not occur if two particles go on-shell in the same loop. This statement would require further analysis in general as it has been investigated in detail only in Ref. \cite{Meuren_2015} in the case of the one-loop polarization operator (see Fig. \ref{FD_PO}).
In that work, in fact, it was shown that the recombination (recollision) of the electron and the positron in the loop produced via Breit-Wheeler pair production cannot occur at an arbitrary phase of the laser field but only at specific phases correlated with the phase at which the pair was previously created, such that recollision effects do not feature a pulse-length enhancement. Correspondingly, higher-order vertex corrections are expected not to feature accumulation effects like those arising from the one-particle reducible contributions described by the Schwinger-Dyson equations (\ref{SD_psi})-(\ref{SD_a}), which is also physically intuitive as vertex corrections are local corrections, unrelated to the macroscopic propagation of particles inside the plane wave. This can be explicitly verified in the one-loop vertex correction (see Fig. \ref{FD_VC}) recently computed in Ref. \cite{Di_Piazza_2020_b}.

\begin{figure}
\begin{center}
\includegraphics[width=0.6\columnwidth]{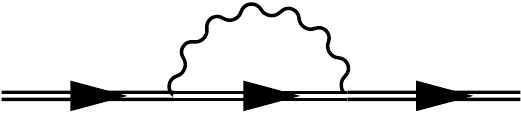}
\caption{The one-loop mass operator in an intense plane wave. The double lines represent exact electron states and propagator in the plane wave (Volkov states and propagator, respectively) \cite{Landau_b_4_1982}.}
\label{FD_MO}
\end{center}
\end{figure}

The Schwinger-Dyson equations (\ref{SD_psi})-(\ref{SD_a}) are clearly impossible to be solved exactly already because it is impossible to compute exactly the mass operator $M(x,y)$ and the polarization operator $P^{\mu\nu}(x,y)$. However, the one-loop mass operator in an arbitrary plane wave (see Fig. \ref{FD_MO}) and the one-loop polarization operator in an arbitrary plane wave (see Fig. \ref{FD_PO}) have been computed in Ref. \cite{Baier_1976_a} and in Refs. \cite{Becker_1975,Baier_1976_b,Meuren_2013}, respectively.
From the perspective of the mentioned accumulation effects, it is important to point out that the inclusion of higher-loop corrections to the mass operator and to the polarization operator would lead to subdominant contributions scaling with higher powers of the fine-structure constant $\alpha$ (we do not consider here the so-called fully non-perturbative regime of strong-field QED at $\chi_0\sim 1/\alpha^{3/2}\gg 1$ and $\xi_0^3\gg \chi_0$, where, according to the so-called Ritus-Narozhny conjecture, the perturbative approach to strong-field QED breaks down \cite{Ritus_1970,Narozhny_1979,Narozhny_1980,Morozov_1981,Akhmedov_1983,Akhmedov_2011,Fedotov_2017,
Podszus_2019,Ilderton_2019,Mironov_2020}).

The conclusion of the above discussion is that, if one would like to take into account accumulation effects depending on the laser pulse duration in nonlinear Compton scattering and in nonlinear Breit-Wheeler pair production but still neglect corrections scaling only with $\alpha$, one could solve the Schwinger-Dyson equations with the one-loop mass operator and polarization operator, find the corresponding electron/positron and photon states including the decaying effects, and use these states to compute the probabilities as the modulus square of the single-vertex amplitude. This is the aim of the present work and we use the electron/positron and photon states determined from the Schwinger-Dyson equations in Ref. \cite{Meuren_2011} and in Ref. \cite{Meuren_2015} within the LCFA, respectively (see Refs. \cite{Dinu_2014,Villalba-Chavez_2016} for a solution of the Schwinger-Dyson equation of the photon field in a plane wave, where radiative effects are treated perturbatively). For the sake of completeness we also present an equivalent but alternative solution of the Schwinger-Dyson equation (\ref{SD_psi}) of the electron field as compared to that in Ref. \cite{Meuren_2011} and we provide more details about the derivation of the solution of the Schwinger-Dyson equation (\ref{SD_a}) of the photon field as given in Ref. \cite{Meuren_2015}. As we will see, these solutions and, in general, the Schwinger-Dyson equations (\ref{SD_psi})-(\ref{SD_a}) apply for electron and photons in-states, respectively. Thus, we also derive the Schwinger-Dyson equations for the positron states and for out-states, and we provide the corresponding solutions under the same conditions as for the electron in-states. Finally, we use these in- and out-states to derive analytical expressions of the probabilities of nonlinear Compton scattering and nonlinear Breit-Wheeler pair production, which feature exponential damping terms describing the decay of the particles in the plane wave.

\section{Basic definitions and notation}

We consider a plane-wave background field described by the four-vector potential $A^{\mu}(\phi)$, which only depends on the light-cone time $\phi=t-\bm{n}\cdot \bm{x}$. Here, the unit vector $\bm{n}$ defines the propagation direction of the plane wave and can be used to introduce the two four-dimensional quantities $n^{\mu}=(1,\bm{n})$ and $\tilde{n}^{\mu}=(1,-\bm{n})/2$ (note that $\phi=(nx)$). The four-vector potential $A^{\mu}(\phi)$ is a solution of the free wave equation $\partial_{\mu}\partial^{\mu} A^{\nu}=0$ and it is assumed to fulfill the Lorenz-gauge condition $\partial_{\mu}A^{\mu}=0$, with the additional constraint $A^0(\phi)=0$. Thus, if we represent $A^{\mu}(\phi)$ in the form $A^{\mu}(\phi)=(0,\bm{A}(\phi))$, then the Lorenz-gauge condition implies that $\bm{n}\cdot\bm{A}'(\phi)=0$, with the prime in a function of $\phi$ indicating its derivative. If we make the additional assumption that $\bm{A}(\phi)$ vanishes for $\phi\to\pm\infty$, then it is $\bm{n}\cdot\bm{A}(\phi)=0$. By introducing two four-vectors $a_j^{\mu}=(0,\bm{a}_j)$, with $j=1,2$, such that $(na_j)=-\bm{n}\cdot\bm{a}_j=0$ and $(a_ja_{j'})=-\bm{a}_j\cdot\bm{a}_{j'}=-\delta_{jj'}$, with $j,j'=1,2$, the vector potential $\bm{A}(\phi)$ can then be written as $\bm{A}(\phi)=\psi_1(\phi)\bm{a}_1+\psi_2(\phi)\bm{a}_2$, where the two functions $\psi_j(\phi)$ are arbitrary, provided that they vanish for $\phi\to\pm\infty$ and they feature obvious differential properties. The field tensor $F^{\mu\nu}(\phi)=\partial^{\mu}A^{\nu}(\phi)-\partial^{\nu}A^{\mu}(\phi)$ of the plane wave is given by $F^{\mu\nu}(\phi)=n^{\mu}A^{\prime\,\nu}(\phi)-n^{\nu}A^{\prime\,\mu}(\phi)$. 

Since the four-vector potential of the plane-wave field will always be multiplied by the electron charge, it is convenient to introduce the four-vector $\mathcal{A}^{\mu}(\phi)=eA^{\mu}(\phi)$ and the tensor $\mathcal{F}^{\mu\nu}(\phi)=eF^{\mu\nu}(\phi)$. Also, we will consider below only the case of linear polarization along the direction $\bm{a}_1$ and we set $\psi_1(\phi)=A_0\psi(\phi)$, with $A_0>0$ describing the amplitude of the vector potential of the plane-wave field, and $\psi_2(\phi)=0$. In this way, the electromagnetic field tensor $F^{\mu\nu}(\phi)$ can be written as $F^{\mu\nu}(\phi)=F_0^{\mu\nu}\psi'(\phi)$, with $F_0^{\mu\nu}=A_0(n^{\mu}a_1^{\nu}-n^{\nu}a_1^{\mu})$. Also, we introduce for future convenience the dual field $\tilde{F}^{\mu\nu}(\phi)=\tilde{F}^{\mu\nu}_0\psi'(\phi)$, where $\tilde{F}^{\mu\nu}_0=(1/2)\varepsilon^{\mu\nu\lambda\rho}F_{0,\lambda\rho}$, with $\varepsilon^{\mu\nu\lambda\rho}$ being the four-dimensional anti-symmetric tensor and $\varepsilon^{0123}=+1$. Analogous definitions hold for the quantities multiplied by $e$.

The four-dimensional quantities $n^{\mu}$, $\tilde{n}^{\mu}$, and $a^{\mu}_j$ fulfill the relation $\eta^{\mu\nu}=n^{\mu}\tilde{n}^{\nu}+\tilde{n}^{\mu}n^{\nu}-a_1^{\mu}a_1^{\nu}-a_2^{\mu}a_2^{\nu}$ (note that $(n\tilde{n})=1$ and $(\tilde{n}a_j)=0$). Below, we will refer to the transverse ($\perp$) plane as the plane spanned by the two perpendicular unit vectors $\bm{a}_j$. Thus, together with the light-cone time $\phi=t-\bm{n}\cdot \bm{x}$, we also introduce the remaining three light-cone coordinates $T=(\tilde{n}x)=(t+\bm{n}\cdot \bm{x})/2$, and $\bm{x}_{\perp}=(x_{\perp,1},x_{\perp,2})=-((xa_1),(xa_2))=(\bm{x}\cdot\bm{a}_1,\bm{x}\cdot\bm{a}_2)$. Analogously, the light-cone coordinates of an arbitrary four-vector $v^{\mu}=(v_0,\bm{v})$ will be indicated as $v_-=(nv)=v_0-\bm{n}\cdot \bm{v}$, $v_+=(\tilde{n}v)=(v_0+\bm{n}\cdot \bm{v})/2$, and $\bm{v}_{\perp}=(v_{\perp,1},v_{\perp,2})=-((va_1),(va_2))=(\bm{v}\cdot\bm{a}_1,\bm{v}\cdot\bm{a}_2)$. 

The Volkov states are the exact solutions of the Dirac equation in a plane wave \cite{Volkov_1935,Landau_b_4_1982}. Below, the four-vector $p^{\mu}=(\varepsilon,\bm{p})$ indicates an on-shell electron four-momentum, i.e., $\varepsilon =\sqrt{m^2+\bm{p}^2}$. The positive-energy Volkov states $U^{(\text{in}/\text{out})}_s(p,x)$ can be classified by means of the asymptotic momentum quantum numbers $\bm{p}$ and of the asymptotic spin quantum number $s=1,2$ at $\phi\to \mp\infty$. Since below we will also consider off-shell four-momenta, for notational simplicity, we indicate the functional dependence on the four components of the electron four-momentum $p^{\mu}$, although the energy is a function of $\bm{p}$. The Volkov state $U^{(\text{in}/\text{out})}_s(p,x)$ can be written as $U^{(\text{in}/\text{out})}_s(p,x)=e^{i\Phi^{(\text{in}/\text{out})}(p)}E(p,x)u_s(p)$, where
\begin{equation}
\Phi^{(\text{in}/\text{out})}(p)=-\int_{\mp\infty}^0d\varphi\left[\frac{(p\mathcal{A}(\varphi))}{p_-}-\frac{\mathcal{A}^2(\varphi)}{2p_-}\right],
\end{equation}
where
\begin{equation}
\label{E_p}
E(p,x)=\bigg[1+\frac{\hat{n}\hat{\mathcal{A}}(\phi)}{2p_-}\bigg]e^{i\left\{-(px)-\int_0^{\phi}d\varphi\left[\frac{(p\mathcal{A}(\varphi))}{p_-}-\frac{\mathcal{A}^2(\varphi)}{2p_-}\right]\right\}},
\end{equation}
and where $u_s(p)$ is the free, positive-energy spinor normalized as $u^{\dag}_s(p)u_{s'}(p)=2\varepsilon \delta_{ss'}$ \cite{Landau_b_4_1982}. In the expression of the Volkov in- and out-states we have explicitly indicated a physically inconsequential overall phase $\Phi^{(\text{in}/\text{out})}(p)$ for future convenience and in Eq. (\ref{E_p}) we have introduced the notation $\hat{v}=\gamma^{\mu}v_{\mu}$ for a generic four-vector $v^{\mu}$ \cite{Landau_b_4_1982}. Also, the spin quantization axis is conveniently chosen along the magnetic field of the plane wave in the rest frame of the electron, i.e., the spin four-vector $s^{\mu}$ is given by $s^{\mu}=\tilde{F}_0^{\mu\nu}p_{\nu}/(p_-A_0)$, with $s^2=-1$, and $\gamma^5\hat{s}u_s(p)=su_s(p)$, where $\gamma^5=i\gamma^0\gamma^1\gamma^2\gamma^3$. Analogously, we introduce the negative-energy Volkov states $V^{(\text{in}/\text{out})}_s(p,x)=e^{i\Phi^{(\text{in}/\text{out})}(-p)}E(-p,x)v_s(p)$, with $v_s(p)$ being the free, negative-energy spinor normalized as $v^{\dag}_s(p)v_{s'}(p)=2\varepsilon \delta_{ss'}$ \cite{Landau_b_4_1982}.

The expression in Eq. (\ref{E_p}) can also formally be used for the matrix $E(l,x)$, where $l^{\mu}=(l^0,\bm{l})$ is a generic off-shell four-momentum, and this matrix fulfills the identities \cite{Ritus_1985,Di_Piazza_2018_d}:
\begin{align}
\label{Compl_x}
\int d^4x \bar{E}(l,x)E(l',x)&=(2\pi)^4\delta^4(l-l'),\\
\label{Compl_l}
\int \frac{d^4l}{(2\pi)^4} E(l,x)\bar{E}(l,y)&=\delta^4(x-y),\\
\label{E_p_Comm}
\gamma^{\mu}[i\partial_{\mu}-\mathcal{A}_{\mu}(\phi)]E(l,x)&=E(l,x)\hat{l},
\end{align}
where $l^{\prime\,\mu}=(l^{\prime\,0},\bm{l}')$ is another off-shell four-momentum and where, for a generic matrix $M$ in the Dirac space, we have introduced the notation $\bar{M}=\gamma^0M^{\dag}\gamma^0$.

Concerning the states of the radiation field, we will indicate as $q^{\mu}=(\omega,\bm{q})$, with $\omega=|\bm{q}|$, the generic on-shell four-momentum of the photon and the two transverse (linear) polarization states are conveniently identified by means of the four-vector $\Lambda_1^{\mu}(q)=F_0^{\mu\nu}q_{\nu}/(q_-A_0)$ and the pseudo-four-vector $\Lambda_2^{\mu}(q)=\tilde{F}_0^{\mu\nu}q_{\nu}/(q_-A_0)$, which fulfill the relations $(\Lambda_j(q)\Lambda_{j'}(q))=-\delta_{jj'}$, with $j,j'=1,2$.

\section{Decaying electron/positron states}
Let us consider the Schwinger-Dyson equation (\ref{SD_psi}) for the electron-positron field, which we rewrite here in the case of the plane wave background field introduced above:
\begin{equation}
\label{SD_psi_L}
\{\gamma^{\mu}[i\partial_{\mu}-\mathcal{A}_{\mu}(\phi)]-m\}\Psi(x)=\int d^4y\, M_L(x,y)\Psi(y),
\end{equation}
where $M_L(x,y)$ is now the mass operator in the plane wave. 

Before starting solving this equation, we notice that, by introducing the Volkov propagator $G_V(x,y)$ (the usual Feynman prescription is understood for avoiding the poles), the solution of Eq. (\ref{SD_psi_L}) can be formally be written as the series
\begin{equation}
\label{Psi_i_0}
\begin{split}
\Psi(x)&=\Psi_V(x)+\int d^4yd^4z\, G_V(x,y)M_L(y,z)\Psi_V(z)\\
&\quad+\int d^4yd^4zd^4rd^4s\, G_V(x,y)M_L(y,z)G_V(z,r)M_L(r,s)\Psi_V(s)+\cdots
\end{split}
\end{equation}
depending on the corresponding (Volkov) solution $\Psi_V(x)$ of the Dirac equation $\{\gamma^{\mu}[i\partial_{\mu}-\mathcal{A}_{\mu}(\phi)]-m\}\Psi_V(x)=0$. By imagining to use the state $\Psi(x)$ to compute a Feynman amplitude,
\begin{figure}
\begin{center}
\includegraphics[width=0.9\columnwidth]{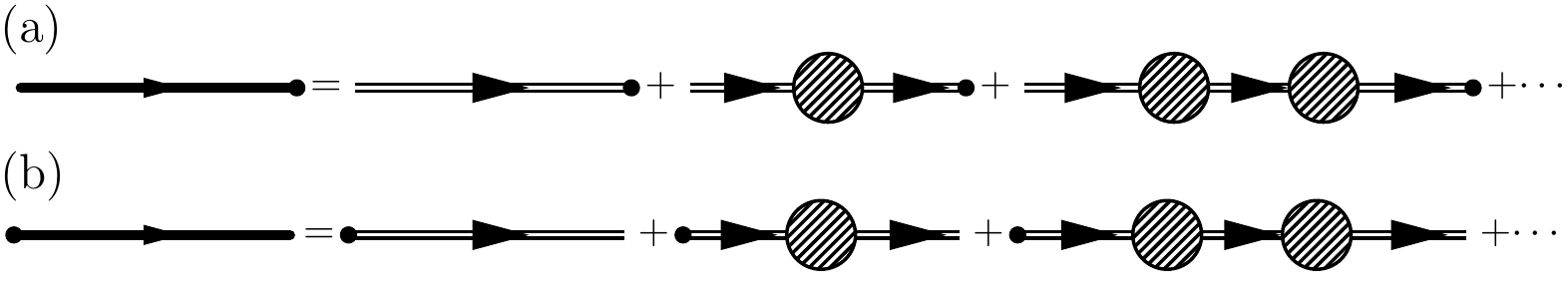}
\caption{The thick lines indicate the exact incoming electron line (a) and the exact outgoing electron line (b), and are symbolically expressed as series expansion in terms of Volkov propagators (double internal lines) and mass operators (shaded circles).}
\label{FD_Exact_E_Line}
\end{center}
\end{figure}
we realize that the solution in Eq. (\ref{Psi_i_0}) is appropriate for an electron in-state $\Psi_e^{(\text{in})}(x)$ (see Fig. \ref{FD_Exact_E_Line}(a)):
\begin{equation}
\label{Psi_i}
\begin{split}
\Psi_e^{(\text{in})}(x)&=\Psi^{(\text{in})}_{e,V}(x)+\int d^4yd^4z\, G_V(x,y)M_L(y,z)\Psi^{(\text{in})}_{e,V}(z)\\
&\quad+\int d^4yd^4zd^4rd^4s\, G_V(x,y)M_L(y,z)G_V(z,r)M_L(r,s)\Psi^{(\text{in})}_{e,V}(s)+\cdots.
\end{split}
\end{equation}
For this reason, we rewrite Eq. (\ref{SD_psi_L}) as
\begin{equation}
\label{SD_psi_L_i}
\{\gamma^{\mu}[i\partial_{\mu}-\mathcal{A}_{\mu}(\phi)]-m\}\Psi_e^{(\text{in})}(x)=\int d^4y\, M_L(x,y)\Psi_e^{(\text{in})}(y).
\end{equation}

Now, for an exact electron out-state $\Psi_e^{(\text{out})}(x)$, one rather needs the series (see Fig. \ref{FD_Exact_E_Line}(b))
\begin{equation}
\label{Psi_b_o}
\begin{split}
\bar{\Psi}_e^{(\text{out})}(x)&=\bar{\Psi}^{(\text{out})}_{e,V}(x)+\int d^4yd^4z\, \bar{\Psi}^{(\text{out})}_{e,V}(z)M_L(z,y)G_V(y,x)\\
&\quad+\int d^4yd^4zd^4rd^4s\, \bar{\Psi}^{(\text{out})}_{e,V}(s)M_L(s,r)G_V(r,z)M_L(z,y)G_V(y,x)+\cdots,
\end{split}
\end{equation}
which is not simply the Dirac conjugated of Eq. (\ref{Psi_i}). Thus, for computing the exact electron out-state $\Psi_e^{(\text{out})}(x)$, we need to solve the Schwinger-Dyson equation
\begin{equation}
\label{SD_psi_L_b_o}
\bar{\Psi}_e^{(\text{out})}(x)\{\gamma^{\mu}[-i\hspace{-0.13cm}\stackrel{\leftarrow}{\partial}_{\mu}-\mathcal{A}_{\mu}(\phi)]-m\}=\int d^4y\, \bar{\Psi}_e^{(\text{out})}(y)M_L(y,x).
\end{equation}
or, equivalently, the equation
\begin{equation}
\label{SD_psi_L_o}
\{\gamma^{\mu}[i\partial_{\mu}-\mathcal{A}_{\mu}(\phi)]-m\}\Psi_e^{(\text{out})}(x)=\int d^4y\, \bar{M}_L(y,x)\Psi_e^{(\text{out})}(y).
\end{equation}
By recalling the Feynman rules for incoming and outgoing positrons, it is easy to derive that the Schwinger-Dyson equations for the states $\Psi_p^{(\text{in})}(x)$ and $\Psi_p^{(\text{out})}(x)$ are
\begin{align}
\label{SD_psi_L_p_i}
\{\gamma^{\mu}[i\partial_{\mu}-\mathcal{A}_{\mu}(\phi)]-m\}\Psi_p^{(\text{in})}(x)&=\int d^4y\, \bar{M}_L(y,x)\Psi_p^{(\text{in})}(y),\\
\label{SD_psi_L_p_o}
\{\gamma^{\mu}[i\partial_{\mu}-\mathcal{A}_{\mu}(\phi)]-m\}\Psi_p^{(\text{out})}(x)&=\int d^4y\, M_L(x,y)\Psi_p^{(\text{out})}(y).
\end{align}

At this point, it is sufficient to outline the derivation of the solution of Eq. (\ref{SD_psi_L_i}) as the other equations (\ref{SD_psi_L_o})-(\ref{SD_psi_L_p_o}) can be solved in an analogous way.

As we have explained in the introduction, it is sufficient for our purposes to consider the one-loop mass operator within the LCFA, which we indicate as $M^{(1)}_L(x,y)$ (see Fig. \ref{FD_MO}). It is convenient first to consider the one-loop mass operator $M^{(1)}_L(l,l')$ in momentum space, defined as
\begin{equation}
M^{(1)}_L(l,l')=\int d^4xd^4y\bar{E}(l,x)M^{(1)}_L(x,y)E(l',y).
\end{equation}
This object can be easily computed from the general expression of the one-loop mass operator in a plane-wave field in Ref. \cite{Baier_1976_a} in the LCFA limit $\xi_0\to\infty$. The resulting expression is given by
\begin{equation}
\label{M_llp}
\begin{split}
M^{(1)}_L(l,l')&=(2\pi)^3\delta^2(\bm{l}_{\perp}-\bm{l}'_{\perp})\delta(l_--l'_-)\int d\phi\, e^{-i(l'_+-l_+)\phi}\frac{\alpha}{2\pi}\int_0^{\infty}\frac{du}{u}\int_0^{\infty}\frac{dv}{(1+v)^2}\\
&\quad\times e^{-i\left[(1+v)\frac{\lambda^2}{m^2}+v^2+v\left(1-\frac{l^{\prime\,2}}{m^2}\right)\right]u}
\bigg\{\bigg(2m-\frac{\hat{l}'}{1+v}\bigg)\left[e^{-\frac{i}{3}v^4\chi_l^2(\phi)u^3}-1\right]+e^{-\frac{i}{3}v^4\chi_l^2(\phi)u^3}\\
&\quad\times\left[\frac{2u^2v^2}{m^4}\left(1+\frac{v}{3}\right)(l\mathcal{F}^2(\phi)\gamma)+i\frac{uv}{m}\sigma_{\mu\nu}\mathcal{F}^{\mu\nu}(\phi)-imuv\frac{2+v}{1+v}\chi_l(\phi)\gamma^5\hat{s}\right]\\
&\quad+\bigg(2m-\frac{\hat{l}'}{1+v}\bigg)\left[1-e^{iuv\left(1-\frac{l^{\prime\,2}}{m^2}\right)}\right]-2iuv\frac{1+2v}{1+v}(\hat{l}'-m)e^{iuv\left(1-\frac{l^{\prime\,2}}{m^2}\right)}\bigg\},
\end{split}
\end{equation}
where $\lambda^2$ is the square of the fictitious photon mass, which has been added for completeness and which will be ultimately set equal to zero because we will only use the mass operator on the mass shell \cite{Ritus_1985}, and where $\sigma^{\mu\nu}=(i/2)[\gamma^{\mu},\gamma^{\nu}]$. As expected, the quantity $M^{(1)}_L(l,l')$ is exactly the Fourier transform in $\phi$ of the corresponding expression of the one-loop mass operator in a constant crossed field found in Ref. \cite{Ritus_1970}, with the replacements $\mathcal{F}_0^{\mu\nu}\to \mathcal{F}^{\mu\nu}(\phi)=\mathcal{F}_0^{\mu\nu}\psi'(\phi)$ and $\chi_0\to \chi_l(\phi)=(l_-/p_-)\chi(\phi)$, with $\chi(\phi)=(p_-/m)A_0\psi'(\phi)/F_{cr}$. As a technical remark, we observe that the quantity $M^{(1)}_L(l,l')$ in Eq. (\ref{M_llp}) is the renormalized expression of the one-loop mass operator, the renormalization being carried out as in vacuum \cite{Ritus_1970,Baier_1976_a}.

Going back to configuration space by using Eqs. (\ref{Compl_x})-(\ref{Compl_l}), we obtain
\begin{equation}
\label{M_xy}
M^{(1)}_L(x,y)=\int \frac{d^4l}{(2\pi)^4}\frac{d^4l'}{(2\pi)^4} E(l,x)M^{(1)}_L(l,l')\bar{E}(l',y)=\int \frac{d^4l}{(2\pi)^4}E(l,x)M_L^{(1)}(l,\phi_x)\bar{E}(l,y),
\end{equation}
where
\begin{equation}
\label{M}
\begin{split}
M_L^{(1)}(l,\phi)&=\frac{\alpha}{2\pi}\int_0^{\infty}\frac{du}{u}\int_0^{\infty}\frac{dv}{(1+v)^2} e^{-i\left[(1+v)\frac{\lambda^2}{m^2}+v^2+v\left(1-\frac{l^2}{m^2}\right)\right]u}\\
&\quad\times
\bigg\{\bigg(2m-\frac{\hat{l}}{1+v}\bigg)\left[e^{-\frac{i}{3}v^4\chi_l^2(\phi)u^3}-1\right]+e^{-\frac{i}{3}v^4\chi_l^2(\phi)u^3}\\
&\quad\times\left[\frac{2u^2v^2}{m^4}\left(1+\frac{v}{3}\right)(l\mathcal{F}^2(\phi)\gamma)+i\frac{uv}{m}\sigma_{\mu\nu}\mathcal{F}^{\mu\nu}(\phi)-imuv\frac{2+v}{1+v}\chi_l(\phi)\gamma^5\hat{s}\right]\\
&\quad+\bigg(2m-\frac{\hat{l}}{1+v}\bigg)\left[1-e^{iuv\left(1-\frac{l^2}{m^2}\right)}\right]-2iuv\frac{1+2v}{1+v}(\hat{l}-m)e^{iuv\left(1-\frac{l^2}{m^2}\right)}\bigg\},
\end{split}
\end{equation}
and where $\phi_x$ is the minus light-cone coordinate of the spacetime point $x$.

Two observations are in order about this intermediate step. First, in the second equality in Eq. (\ref{M_xy}) one exploits the delta functions in Eq. (\ref{M_llp}) to take the integrals over $d^2\bm{l}'_{\perp}$ and $dl'_-$ such that the integral over $d^4l$ is originally given over the variables $\bm{l}_{\perp}$, $l_-$, and $l'_+$. Only after one renames the integration variable $l'_+$ as $l_+$, one can express the result as the four-dimensional integral over $d^4l$. Second, the appearance of the quantity $M_L^{(1)}(l,\phi_x)$ evaluated at $\phi_x$ looks ``asymmetric''. However, it arises because we have decided to express the mass operator in Eq. (\ref{M_llp}) in terms of $l^{\prime\,2}$ and $\hat{l}'$. It is easy to show that one can perform a shift in the variable $\phi$ in Eq. (\ref{M_llp}) and obtain an equivalent expression within the LCFA only in terms of $l^2$ and $\hat{l}$. This would result in an equivalent expression (within the LCFA) of $M^{(1)}_L(x,y)$, where the quantity $M_L^{(1)}(l,\phi_y)$ appears on the right-hand side of Eq. (\ref{M_xy}), with $\phi_y$ being the minus light-cone coordinate of the spacetime point $y$. The expression featuring the quantity $M_L^{(1)}(l,\phi_x)$ has been chosen for later convenience.

Now, similarly as in Ref. \cite{Meuren_2011} and without loss of generality, we decide to determine how the effects of the mass operator in the right-hand side of the Schwinger-Dyson equation (\ref{SD_psi_L_i}) modify the Volkov in-state $U^{(\text{in})}_s(p,x)$ (recall that $p^{\mu}=(\varepsilon,\bm{p})$, with $\varepsilon=\sqrt{m^2+\bm{p}^2}$). Thus, we seek for a solution of Eq. (\ref{SD_psi_L_i}) of the form
\begin{equation}
\Psi_e^{(\text{in})}(x)=e^{i\Phi^{(\text{in})}(p)}f^{(\text{in})}_s(p,\phi)E(p,x)u_s(p),
\end{equation}
where $f^{(\text{in})}_s(p,\phi)$ is a function to be determined, which satisfies the initial condition $\lim_{\phi\to-\infty}f^{(\text{in})}_s(p,\phi)=1$. This initial condition corresponds to the physical requirement that the electron state coincides with the free state $\exp(-i(px))u_s(p)$ before the electron interacts with the plane-wave field. By replacing the above expression of $\Psi_e^{(\text{in})}(x)$ in Eq. (\ref{SD_psi_L}), by exploiting the relation in Eq. (\ref{E_p_Comm}), and by multiplying the resulting expression by $\bar{u}_{s'}(p)\bar{E}(p,x)$, we obtain
\begin{equation}
\label{SD_f_0}
\begin{split}
&2ip_-\delta_{ss'}\frac{df^{(\text{in})}_s(p,\phi_x)}{d\phi_x}\\
&\quad=\int d^4y\frac{d^4l}{(2\pi)^4}\bar{u}_{s'}(p)\bar{E}(p,x)E(l,x)M_L^{(1)}(l,\phi_x)\bar{E}(l,y)E(p,y)u_s(p)f^{(\text{in})}_s(p,\phi_y),
\end{split}
\end{equation}
where we have used the relation $\bar{u}_{s'}(p)\hat{n}u_s(p)=2p_-\delta_{ss'}$.

Now, the integrals in $\bm{y}_{\perp}=-((ya_1),(ya_2))$ and in $T_y=(y_0+\bm{n}\cdot\bm{y})/2$ in Eq. (\ref{SD_f_0}) can be taken analytically because the matrices $E(p,x)$ depend on the transverse and the plus light-cone coordinates only linearly in the phase. Thus, these integrals provide delta functions enforcing that $\bm{l}_{\perp}=\bm{p}_{\perp}$ and $l_-=p_-$. By taking the corresponding integrals in $\bm{l}_{\perp}$ and $l_-$, we obtain
\begin{equation}
\label{SD_f}
2ip_-\delta_{ss'}\frac{df^{(\text{in})}_s(p,\phi_x)}{d\phi_x}=\int d\phi_y\frac{dl_+}{2\pi}e^{i(p_+-l_+)(\phi_x-\phi_y)}\bar{u}_{s'}(p)M_L^{(1)}(l,\phi_x)u_s(p)f^{(\text{in})}_s(p,\phi_y).
\end{equation}
Here, the four-momentum $l^{\mu}$ is intended to have light-cone components $p_-$, $\bm{p}_{\perp}$, and $l_+$, and it is important to notice from Eq. (\ref{M}) that $l_+$ only appears linearly in the exponents via the squared four-momentum $l^2=2p_-l_+-\bm{p}_{\perp}^2$ and also linearly in the pre-exponential function via the matrix $\hat{l}=l_+\hat{n}+p_-\hat{\tilde{n}}-(pa_1)\hat{a}_1-(pa_2)\hat{a}_2$. This observation allows us to take the integral in $l_+$ analytically. It is first convenient to perform the change of variable from $l_+$ to $r_+=l_+-p_+$. The result of the integral over $r_+$ is rather cumbersome but it can be simplified by means of two remarks related to the dependence of $M_L^{(1)}(l,\phi_x)$ on $l_+$ and then on $r_+$ [see Eq. (\ref{M})]:
\begin{enumerate}
\item The presence of the exponential functions $\exp(2iuvp_-r_+/m^2)$ leads to the presence of delta-function $\delta(\phi_y-\phi_x+2uvp_-/m^2)$, which result in the function $f^{(\text{in})}_s(p,\phi_y)$ to be computed at $\phi_y=\phi_x-2uvp_-/m^2$ for those terms. However, within the LCFA and since the function $f^{(\text{in})}_s(p,\phi_y)$ can be \textit{a posteriori} ascertained to be sufficiently smooth in $\phi_y$, one can approximate $f^{(\text{in})}_s(p,\phi_x-2uvp_-/m^2)\approx f^{(\text{in})}_s(p,\phi_x)$ (see also below).
\item The terms proportional to $r_+\hat{n}$ in the pre-exponent turn into terms containing the derivative of the function $f_s(p,\phi_y)$ also on the right-hand side of Eq. (\ref{SD_f}). However, these terms can be easily proven to have exactly the same structure of the left-hand side of Eq. (\ref{SD_f}), apart of course, being already proportional to $\alpha$. Thus, after exploiting the delta function in the light-cone times, by imagining to combine these terms with the left-hand side of Eq. (\ref{SD_f}) and to divide the resulting equation by the overall coefficient of $df^{(\text{in})}_s(p,\phi_x)/d\phi_x$, one finally concludes that these additional terms lead to higher-order corrections in $\alpha$ and can be ignored within our leading-order analysis.
\end{enumerate}
By means of these considerations, it is straightforward to obtain that
\begin{equation}
\label{SD_f_1}
ip_-\delta_{ss'}\frac{df^{(\text{in})}_s(p,\phi)}{d\phi}=mM_{ss'}(p,\phi)f^{(\text{in})}_s(p,\phi),
\end{equation}
where $M_{ss'}(p,\phi)=\bar{u}_{s'}(p)M(p,\phi)u_s(p)/\bar{u}_s(p)u_s(p)$, with
\begin{equation}
\begin{split}
M(p,\phi)&=\frac{\alpha}{2\pi}\int_0^{\infty}\frac{du}{u}\int_0^{\infty}\frac{dv}{(1+v)^2}e^{-iv^2u}
\left\{\left(2m-\frac{m}{1+v}\right)\left[e^{-\frac{i}{3}v^4\chi^2(\phi)u^3}-1\right]+e^{-\frac{i}{3}v^4\chi^2(\phi)u^3}\right.\\
&\left.\quad\times\left[\frac{2u^2v^2}{m^4}\left(1+\frac{v}{3}\right)(p\mathcal{F}^2(\phi)\gamma)+i\frac{uv}{m}\sigma_{\mu\nu}\mathcal{F}^{\mu\nu}(\phi)-imuv\frac{2+v}{1+v}\chi(\phi)\gamma^5\hat{s}\right]\right\}.
\end{split}
\end{equation}
It is worth noticing that the quantity $M(p,\phi)$ does not contain vacuum terms and it vanishes if the plane wave vanishes. This is in agreement with the well-known fact that on-shell states do not undergo radiative corrections in vacuum \cite{Landau_b_4_1982}.

At this point, by using the properties of the free states $u_s(p)$, it is not difficult to prove that the matrix $M_{ss'}(p,\phi)$ is diagonal. Thus, we conclude that the function $f^{(\text{in})}_s(p,\phi)$ satisfies the equation
\begin{equation}
\label{Eq_f_f}
i\frac{df^{(\text{in})}_s(p,\phi)}{d\phi}=\frac{m}{p_-}M_s(p,\phi)f^{(\text{in})}_s(p,\phi),
\end{equation}
where
\begin{equation}
\begin{split}
M_s(p,\phi)&=m\frac{\alpha}{2\pi}\int_0^{\infty}\frac{du}{u}\int_0^{\infty}\frac{dv}{(1+v)^2}e^{-iv^2u}
\left\{\frac{1+2v}{1+v}\left[e^{-\frac{i}{3}v^4\chi^2(\phi)u^3}-1\right]\right.\\
&\quad\left. +e^{-\frac{i}{3}v^4\chi^2(\phi)u^3}\left[2u^2v^2\left(1+\frac{v}{3}\right)\chi^2(\phi)+is\frac{uv^2}{1+v}\chi(\phi)\right]\right\}.
\end{split}
\end{equation}
The quantity $M_s(p,\phi)$ is easily shown, by means of changes of variables in $u$ and $v$ in the first term, to exactly coincide with the spin-dependent mass correction in a constant crossed field, with the replacement $\chi_0\to \chi(\phi)$ \cite{Ritus_1970,Baier_1971}:
\begin{equation}
\label{M_s}
M_s(p,\phi)=m\frac{\alpha}{2\pi}\int_0^{\infty} du\int_0^{\infty}\frac{dv}{(1+v)^3}e^{-iu\left[1+\frac{1}{3}\frac{\chi^2(\phi)}{v^2}u^2\right]}\left[\frac{5+7v+5v^2}{3}\frac{\chi^2(\phi)}{v^2}u+is\chi(\phi)\right].
\end{equation}

Now, Eq. (\ref{Eq_f_f}) can be easily integrated. In fact, by imposing the initial condition $\lim_{\phi\to-\infty}f^{(\text{in})}_s(p,\phi)=1$, we obtain that the radiatively-corrected positive-energy Volkov in-state $U^{(\text{in})}_{R,s}(p,x)$, which include the decay of the state itself, is given by
\begin{equation}
\label{U_i}
\begin{split}
U^{(\text{in})}_{R,s}(p,x)&=e^{i\Phi^{(\text{in})}(p)}e^{-i\frac{m}{p_-}\int_{-\infty}^{\phi}d\varphi M_s(p,\varphi)}E(p,x)u_s(p)\\
&=\bigg[1+\frac{\hat{n}\hat{\mathcal{A}}(\phi)}{2p_-}\bigg]e^{i\left\{-(px)-\int_{-\infty}^{\phi}d\varphi\left[\frac{(p\mathcal{A}(\varphi))}{p_-}-\frac{\mathcal{A}^2(\varphi)}{2p_-}+\frac{m}{p_-}M_s(p,\varphi)\right]\right\}}u_s(p).
\end{split}
\end{equation}
Within the LCFA this expression exactly coincides with the corresponding solution found in Ref. \cite{Meuren_2011} (the additional term in the pre-exponential function in Ref. \cite{Meuren_2011} can be shown to be equivalent to one having the same matrix structure of the field-dependent term of the Volkov state and being smaller than that term by a factor proportional to $\xi_0\gg 1$, which can be ignored at the leading order in the LCFA considered here). Also, we observe in relation to the point 1. below Eq. (\ref{SD_f}), that a shift of $\phi$ in the decaying exponential function by a term scaling as $p_-/m^2$ would result into a correction of the order of $\alpha$ to the pre-exponential function, which can be neglected under our approximations.

Analogously, by solving the Schwinger-Dyson equation (\ref{SD_psi_L_p_i}), one obtains that the positron state $V^{(\text{in})}_s(p,x)$ turns into the radiatively-corrected state
\begin{equation}
\label{V_i}
V^{(\text{in})}_{R,s}(p,x)=\bigg[1-\frac{\hat{n}\hat{\mathcal{A}}(\phi)}{2p_-}\bigg]e^{i\left\{(px)-\int_{-\infty}^{\phi}d\varphi\left[\frac{(p\mathcal{A}(\varphi))}{p_-}+\frac{\mathcal{A}^2(\varphi)}{2p_-}-\frac{m}{p_-}M^*_s(-p,\varphi)\right]\right\}}v_s(p),
\end{equation}
which includes the decay effects. Note that $M_s(-p,\varphi)=M_{-s}(p,\varphi)$ [see Eq. (\ref{M_s})], which correctly corresponds to the interaction magnetic energy of the positron intrinsic magnetic moment having the opposite sign of that of the electron.

Note that the radiatively-corrected electron in-states are normalized as
\begin{align}
\frac{\bar{U}^{(\text{in})}_{R,s}(p,x)U^{(\text{in})}_{R,s}(p,x)}{\bar{u}_s(p)u_s(p)}&=e^{\frac{2m}{p_-}\int_{-\infty}^{\phi}d\varphi\, \text{Im}[M_s(p,\varphi)]},\\
\frac{\bar{V}^{(\text{in})}_{R,s}(p,x)V^{(\text{in})}_{R,s}(p,x)}{\bar{v}_s(p)v_s(p)}&=e^{\frac{2m}{p_-}\int_{-\infty}^{\phi}d\varphi\, \text{Im}[M_s(-p,\varphi)]}.
\end{align}
By using the optical theorem, it can be shown that the quantity $-(2m/p_-)\text{Im}[M_s(\pm p,\phi)]$ is the total probability per unit of light-cone time $\phi$ that an electron/positron with initial four-momentum $p^{\mu}$ and spin quantum number $s$ emits a photon \cite{Baier_1976_a}. Then, these normalization conditions exactly describe the decay of the state as due to the fact that electron/positron states are unstable under emission of photons in the presence of the plane wave.

Finally, by solving in a completely analogous way the Schwinger-Dyson equations (\ref{SD_psi_L_o}) and (\ref{SD_psi_L_p_o}), one obtains the following expressions for the radiatively-corrected electron and positron out-states:
\begin{align}
\label{U_o}
U^{(\text{out})}_{R,s}(p,x)&=\bigg[1+\frac{\hat{n}\hat{\mathcal{A}}(\phi)}{2p_-}\bigg]e^{i\left\{-(px)+\int_{\phi}^{\infty}d\varphi\left[\frac{(p\mathcal{A}(\varphi))}{p_-}-\frac{\mathcal{A}^2(\varphi)}{2p_-}+\frac{m}{p_-}M^*_s(p,\varphi)\right]\right\}}u_s(p),\\
\label{V_o}
V^{(\text{out})}_{R,s}(p,x)&=\bigg[1-\frac{\hat{n}\hat{\mathcal{A}}(\phi)}{2p_-}\bigg]e^{i\left\{(px)+\int_{\phi}^{\infty}d\varphi\left[\frac{(p\mathcal{A}(\varphi))}{p_-}+\frac{\mathcal{A}^2(\varphi)}{2p_-}-\frac{m}{p_-}M_s(-p,\varphi)\right]\right\}}v_s(p).
\end{align}
As expected and appropriate for out-states, in this case the decaying exponential functions feature light-cone integrals from $\phi$ to $\infty$.

\section{Decaying photon states}

\begin{figure}
\begin{center}
\includegraphics[width=0.9\columnwidth]{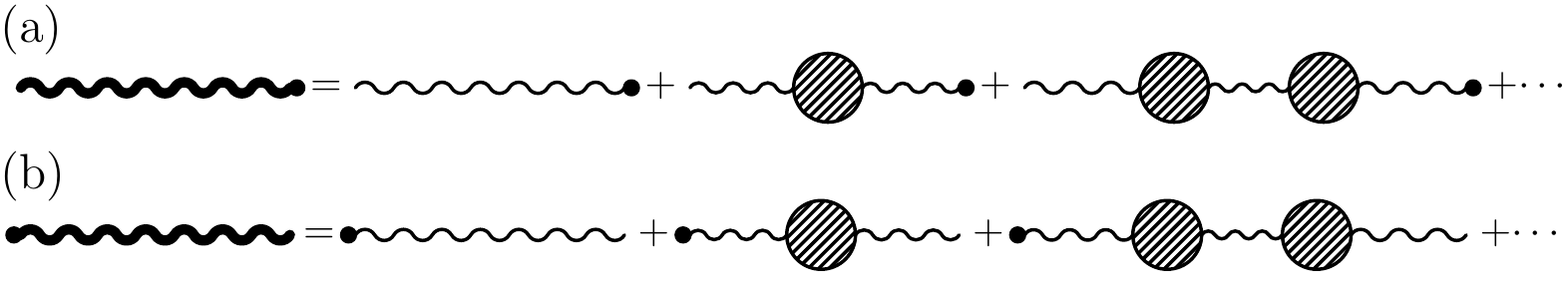}
\caption{The thick wiggly lines indicate the exact incoming photon line (a) and the exact outgoing photon line (b), and are symbolically expressed as series expansion in terms of free photon propagator (thin wiggly internal lines) and polarization operators (shaded circles).}
\label{FD_Exact_G_Line}
\end{center}
\end{figure}

Analogously as in the previous section, we first observe that the Schwinger-Dyson equation (\ref{SD_a}) applies to photon in-states $\mathscr{A}^{(\text{in})}_{\nu}(x)$ and, in the case of a background plane wave, can be rewritten as (see Fig. \ref{FD_Exact_G_Line}(a))
\begin{equation}
\label{SD_a_L_i}
-\partial_{\mu}\partial^{\mu}\mathscr{A}^{(\text{in})}_{\nu}(x)=\int d^4y\, P_{L,\nu}^{\quad\lambda}(x,y)\mathscr{A}^{(\text{in})}_{\lambda}(y),
\end{equation}
where $P_L^{\nu\lambda}(x,y)$ is the polarization operator in the plane-wave field. The corresponding Schwinger-Dyson equation for the photon out-state $\mathscr{A}^{(\text{out})}_{\nu}(x)$ is given by (see Fig. \ref{FD_Exact_G_Line}(b))
\begin{equation}
\label{SD_a_L_o}
-\partial_{\mu}\partial^{\mu}\mathscr{A}^{(\text{out})}_{\nu}(x)=\int d^4y\, P_{L\;\;\nu}^{*\,\lambda}(y,x)\mathscr{A}^{(\text{out})}_{\lambda}(y),
\end{equation}

By limiting to the case of photon in-states, we recall that Eq. (\ref{SD_a_L_i})  has been solved in Ref. \cite{Meuren_2015} for $P_L^{\nu\lambda}(x,y)$ being replaced by the corresponding one-loop polarization operator within the LCFA. Here, for the sake of completeness, we only present a few steps of the derivation in Ref. \cite{Meuren_2015}.

As we have mentioned in the introduction, the one-loop polarization operator, denoted here as $P_L^{(1)\,\mu\nu}(x,y)$, was computed in Refs. \cite{Baier_1976_b,Becker_1975,Meuren_2015}. By limiting to the contributions corresponding to photons with transverse polarization, the polarization operator $P_L^{(1)\,\mu\nu}(x,y)$ within the LCFA can be written in the form \cite{Meuren_2015}
\begin{equation}
\begin{split}
P_L^{(1)\,\mu\nu}(x,y)=\int \frac{d^4 l}{(2\pi)^4}\frac{d^4 l'}{(2\pi)^4} e^{-i (lx)} P_L^{(1)\,\mu\nu}(l,l') e^{i(l'y)},
\end{split}
\end{equation}
where
\begin{equation}
\label{P_llp}
\begin{split}
P_L^{(1)\,\mu\nu}(l,l')&=-(2\pi)^3\delta^2(\bm{l}_{\perp}-\bm{l}'_{\perp})\delta(l_--l'_-)\int d\phi\, e^{-i(l'_+-l_+)\phi}\frac{\alpha}{24\pi}m^2\chi_l^2(\phi) \\
&\quad\times\int_0^{\infty}du\,u\int_0^1 dv (1-v^2)e^{-iu\left[1-\frac{l^{\prime\,2}}{m^2}\frac{1-v^2}{4}+\frac{(1-v^2)^2}{48}\chi_l^2(\phi)u^2\right]}\\
&\quad\times[(3+v^2)\Lambda_1^\mu(l) \Lambda_1^\nu(l)+(6-2v^2)\Lambda_2^\mu(l) \Lambda_2^\nu(l)]
\end{split}
\end{equation}
is the polarization operator in momentum space. Note that $P_L^{(1)\,\mu\nu}(l,l')$ vanishes for a vanishing plane wave (as it is known, the polarization operator in vacuum is proportional to the tensor $l^2\eta^{\mu\nu}-l^{\mu}l^{\nu}$ \cite{Landau_b_4_1982}). Note that the above expression of the one-loop polarization operator is exactly the Fourier transform in the light-cone time $\phi$ of the corresponding expression in a constant crossed field with the replacement $\chi_0\to\chi_l(\phi)$ \cite{Ritus_1970}.

Analogously as in electron/positron case, the polarization operator in configuration space can be written as
\begin{equation}
\label{P_xy}
P_L^{(1)\,\mu\nu}(x,y)=\int \frac{d^4 l}{(2\pi)^4} e^{-i (lx)} P_L^{(1)\,\mu\nu}(l,\phi_x) e^{i(ly)},
\end{equation}
where
\begin{equation}
\label{P_l}
\begin{split}
P_L^{(1)\,\mu\nu}(l,\phi)&=-\frac{\alpha}{24\pi}m^2\chi_l^2(\phi)\int_0^{\infty}du\,u\int_0^1 dv (1-v^2)e^{-iu\left[1-\frac{l^2}{m^2}\frac{1-v^2}{4}+\frac{(1-v^2)^2}{48}\chi_l^2(\phi)u^2\right]}\\
&\quad\times[(3+v^2)\Lambda_1^\mu(l) \Lambda_1^\nu(l)+(6-2v^2)\Lambda_2^\mu(l) \Lambda_2^\nu(l)].
\end{split}
\end{equation}

Now, by using the above expression of the polarization operator, we investigate how radiative corrections represented by the right-hand side of Eq. (\ref{SD_a_L_i}) modify the photon in-state $\mathscr{A}_{j,\mu}^{(\text{in})}(x)=\exp(-i(qx))\Lambda_{j,\mu}(q)$ (recall that $q^{\mu}=(\omega,\bm{q})$, with $\omega=|\bm{q}|$). Thus, we seek for a solution of the Schwinger-Dyson equation (\ref{SD_a_L_i}) of the form
\begin{equation}
\mathscr{A}_{\mu}^{(\text{in})}(x)=g^{(\text{in})}_j(q,\phi)e^{-i(qx)}\Lambda_{j,\mu}(q),
\end{equation}
where the unknown function $g^{(\text{in})}_j(q,\phi)$ has to fulfill the initial condition $\lim_{\phi\to-\infty}g^{(\text{in})}_j(q,\phi)=1$. By replacing this expression in the Schwinger-Dyson equation (\ref{SD_a_L_i}), we obtain
\begin{equation}
2iq_-\delta_{jj'}\frac{dg^{(\text{in})}_j(q,\phi)}{d\phi}=\int d^4y\frac{d^4 l}{(2\pi)^4} e^{-i ((l-q)x)} \Lambda_{j',\mu}(q)P_L^{(1)\,\mu\nu}(l,\phi_x)\Lambda_{j,\nu}(q) e^{i((l-q)y)}g^{(\text{in})}_j(q,\phi_y).
\end{equation}
The integrals over $d^2\bm{y}_{\perp}$ and $dT_y$ can taken analogously as in the electron/positron case and result into the delta functions $\delta^2(\bm{l}_{\perp}-\bm{q}_{\perp})$ and $\delta(l_--q_-)$, respectively, such that the equation for $g^{(\text{in})}_j(q,\phi)$ becomes
\begin{equation}
2iq_-\delta_{jj'}\frac{dg^{(\text{in})}_j(q,\phi)}{d\phi}=\int d\phi_y\frac{dl_+}{2\pi}e^{i(q_+-l_+)(\phi_x-\phi_y)}\Lambda_{j',\mu}(q)P_L^{(1)\,\mu\nu}(l,\phi_x)\Lambda_{j,\nu}(q)g^{(\text{in})}_j(q,\phi_y),
\end{equation}
where now the four-vector $l^{\mu}$ has light-cone components $q_-$, $\bm{q}_{\perp}$, and $l_+$. Due to the special dependence of the quantities $\Lambda^{\mu}_j(q)$ on the four-momentum, one can already ascertain that the right-hand side is also diagonal on the indexes $j$ and $j'$. Also, one can now take the integral over $l_+$ because $P_L^{(1)\,\mu\nu}(l,\phi_x)$ contains $l_+$ only in the exponent via the squared four-momentum $l^2=2q_-l_+-\bm{q}_{\perp}^2$. Exactly with the same reasoning as in the electron/positron case, we can show that the function $g^{(\text{in})}_j(q,\phi)$ has to fulfill the equation
\begin{equation}
iq_-\frac{dg^{(\text{in})}_j(q,\phi)}{d\phi}=mP_j(q,\phi)g^{(\text{in})}_j(q,\phi),
\end{equation}
where
\begin{align}
\label{P_1}
P_1(q,\phi)&=\frac{\alpha}{48\pi}m\chi^2(\phi)\int_0^{\infty}du\,u\int_0^1 dv e^{-iu\left[1+\frac{(1-v^2)^2}{48}\chi^2(\phi)u^2\right]}(1-v^2)(3+v^2),\\
\label{P_2}
P_2(q,\phi)&=\frac{\alpha}{48\pi}m\chi^2(\phi)\int_0^{\infty}du\,u\int_0^1 dv e^{-iu\left[1+\frac{(1-v^2)^2}{48}\chi^2(\phi)u^2\right]}(1-v^2)(6-2v^2).
\end{align}

By accounting for the initial condition on $g^{(\text{in})}_j(q,\phi)$, we finally obtain
\begin{equation}
\label{A_i}
\mathscr{A}_{R,j,\mu}^{(\text{in})}(q,x)=e^{-i(qx)-i\frac{m}{q_-}\int_{-\infty}^{\phi}d\varphi P_j(q,\varphi)}\Lambda_{j,\mu}(q).
\end{equation}
In a completely analogous way, by solving the Schwinger-Dyson equation (\ref{SD_a_L_o}) for the photon out-state, we obtain
\begin{equation}
\label{A_o}
\mathscr{A}_{R,j,\mu}^{(\text{out})}(q,x)=e^{-i(qx)+i\frac{m}{q_-}\int_{\phi}^{\infty}d\varphi P^*_j(q,\varphi)}\Lambda_{j,\mu}(q).
\end{equation}
As in the case of the electron/positron states, one can easily recognize that the damping of the photon states physically corresponds to the fact that photons can decay into electron-positron pairs inside a plane wave. Indeed, from the optical theorem one can show that the quantity $-(2m/q_-)\text{Im}[P_j(q,\phi)]$ corresponds to the total probability per unit of light-cone time $\phi$ that a photon with four-momentum $q^{\mu}$ and polarization $j$ decays into an electron-positron pair \cite{Baier_1976_b}.

\section{Probabilities of nonlinear Compton scattering and nonlinear Breit-Wheeler pair production including the particles states decay}

Having obtained the electron/positron and photon states including the effects of the states decay, it is now straightforward to write down the probabilities of the basic strong-field QED processes at the leading order in $\alpha$ but including the mentioned effects of the decay of the states.

Concerning nonlinear Compton scattering, the leading-order amplitude in $\alpha$ of the process is (we set for simplicity the quantization volume equal to unity)
\begin{equation}
S^{(e^-\to e^-\gamma)}=-ie\int d^4x\,\frac{\bar{U}^{(\text{out})}_{R,s'}(p',x)}{\sqrt{2\varepsilon'}}\frac{\hat{\mathscr{A}}^{(\text{out})\,*}_{R,j}(q,x)}{\sqrt{2\omega}}\frac{U^{(\text{in})}_{R,s}(p,x)}{\sqrt{2\varepsilon}},
\end{equation}
where the meaning of the quantum numbers of the initial and final particles is clear.

After a few standard manipulations, this amplitude can be reduced to a single-dimensional integral over the light-cone time $\phi$:
\begin{equation}
\begin{split}
S^{(e^-\to e^-\gamma)}&=-\frac{ie}{\sqrt{8\varepsilon\varepsilon'\omega}}(2\pi)^3\delta^2(\bm{p}'_{\perp}+\bm{q}_{\perp}-\bm{p}_{\perp})\delta(p'_-+q_--p_-)\\
&\quad\times\int d\phi\,e^{-i\frac{m}{p_-}\int_{-\infty}^{\phi}d\varphi M_s(p,\varphi)-i\int_{\phi}^{\infty}d\varphi \left[\frac{m}{p'_-}M_{s'}(p',\varphi)+\frac{m}{q_-}P_j(q,\varphi)\right]}\\
&\quad\times e^{i\left\{(p'_++q_+-p_+)\phi-\int_{\phi}^{\infty}d\varphi\left[\frac{(p'\mathcal{A}(\varphi))}{p'_-}-\frac{\mathcal{A}^2(\varphi)}{2p'_-}\right]-\int_{-\infty}^{\phi}d\varphi\left[\frac{(p\mathcal{A}(\varphi))}{p_-}-\frac{\mathcal{A}^2(\varphi)}{2p_-}\right]\right\}}\\
&\quad\times\bar{u}_{s'}(p')\left[1-\frac{\hat{n}\hat{\mathcal{A}}(\phi)}{2p'_-}\right]\hat{\Lambda}_j(q)\left[1+\frac{\hat{n}\hat{\mathcal{A}}(\phi)}{2p_-}\right]u_s(p).
\end{split}
\end{equation}
Since the quantities $M_s(p,\phi)$ and $P_j(q,\phi)$ are already computed within the LCFA, the above amplitude is meaningful only within the same approximation, which we implement directly in the probability:
\begin{equation}
\label{P_NCS}
\begin{split}
&P^{(e^-\to e^-\gamma)}=\int \frac{d^3q}{(2\pi)^3}\frac{d^3p'}{(2\pi)^3}|S^{(e^-\to e^-\gamma)}|^2\\
&\quad=\int\frac{d^3q}{16\pi^2}\frac{\alpha}{p_-p'_-\omega}\int d\phi d\phi'e^{-i\frac{m}{p_-}\int_{-\infty}^{\phi}d\varphi M_s(p,\varphi)+i\frac{m}{p_-}\int_{-\infty}^{\phi'}d\varphi M^*_s(p,\varphi)}\\
&\qquad\times e^{-i\int_{\phi}^{\infty}d\varphi \left[\frac{m}{p'_-}M_{s'}(p',\varphi)+\frac{m}{q_-}P_j(q,\varphi)\right]+i\int_{\phi'}^{\infty}d\varphi \left[\frac{m}{p'_-}M^*_{s'}(p',\varphi)+\frac{m}{q_-}P^*_j(q,\varphi)\right]}\\
&\qquad\times e^{i\left\{(p'_++q_+-p_+)(\phi-\phi')+\int_{\phi'}^{\phi}d\varphi\left[\frac{(p'\mathcal{A}(\varphi))}{p'_-}-\frac{\mathcal{A}^2(\varphi)}{2p'_-}-\frac{(p\mathcal{A}(\varphi))}{p_-}+\frac{\mathcal{A}^2(\varphi)}{2p_-}\right]\right\}}\\
&\qquad\times\frac{1}{4}\text{tr}\left\{\left[1-\frac{\hat{n}\hat{\mathcal{A}}(\phi)}{2p'_-}\right]\hat{\Lambda}_j(q)\left[1+\frac{\hat{n}\hat{\mathcal{A}}(\phi)}{2p_-}\right](\hat{p}+m)(1+s\gamma^5\hat{s})\right.\\
&\qquad\left.\times\left[1-\frac{\hat{n}\hat{\mathcal{A}}(\phi')}{2p_-}\right]\hat{\Lambda}_j(q)\left[1+\frac{\hat{n}\hat{\mathcal{A}}(\phi')}{2p'_-}\right](\hat{p}'+m)(1+s'\gamma^5\hat{s}')\right\},
\end{split}
\end{equation}
where we have used the positive-energy electron density matrix $u_s(p)\bar{u}_s(p)=(\hat{p}+m)(1+s\gamma^5\hat{s})/2$ and the analogous for $u_{s'}(p')\bar{u}_{s'}(p')$ \cite{Landau_b_4_1982}. Indeed, by following the procedure as outlined, e.g., in Refs. \cite{Di_Piazza_2018,Di_Piazza_2019}, we can pass from the variables $\phi$ and $\phi'$ to the variables $\phi_+=(\phi+\phi')/2$ and $\phi_-=\phi-\phi'$, and expand the integrand for $\phi_-$ around $\phi_-=0$. Since within the LCFA the phase $\omega_0\phi_-$ can be estimated to be of the order of $1/\xi_0$ and $\xi_0$ is assumed to be much larger than unity, it is appropriate to expand the pre-exponential function up to the linear order in $\phi_-$, the new terms in the phase due to the decay of the states at the zero order in $\phi_-$, as they are already within the LCFA (see also below), and the remaining terms in the phase up to the third order in $\phi_-$, as it is already known.

In this way, we arrive at the following expression of the photon emission probability (see also Refs. \cite{Di_Piazza_2018,Di_Piazza_2019} for the expression of the probability without the decay of the states):
\begin{equation}
\label{P_NCS_f}
\begin{split}
&P^{(e^-\to e^-\gamma)}=\int\frac{d^3q}{16\pi^2}\frac{\alpha}{p_-p'_-\omega}\int d\phi_+e^{2\text{Im}\left\{\frac{m}{p_-}\int_{-\infty}^{\phi_+}d\varphi M_s(p,\varphi)+\int_{\phi_+}^{\infty}d\varphi \left[\frac{m}{p'_-}M_{s'}(p',\varphi)+\frac{m}{q_-}P_j(q,\varphi)\right]\right\}}\\
&\quad\times\int d\phi_- e^{i\frac{m^2}{2p_-}\frac{q_-}{p'_-}\left\{[1+\bm{\pi}_{\perp,e}^2(\phi_+)]\phi_-+\frac{\bm{\mathcal{E}}^2(\phi_+)}{m^2}\frac{\phi_-^3}{12}\right\}}\frac{1}{4}\text{tr}\left\{\left[1-\frac{\hat{n}[\hat{\mathcal{A}}(\phi_+)+\hat{\mathcal{A}}'(\phi_+)\phi_-/2]}{2p'_-}\right]\hat{\Lambda}_j(q)\right.\\
&\quad\times\left[1+\frac{\hat{n}[\hat{\mathcal{\mathcal{A}}}(\phi_+)+\hat{\mathcal{A}}'(\phi_+)\phi_-/2]}{2p_-}\right](\hat{p}+m)(1+s\gamma^5\hat{s})\left[1-\frac{\hat{n}[\hat{\mathcal{A}}(\phi_+)-\hat{\mathcal{A}}'(\phi_+)\phi_-/2]}{2p_-}\right]\\
&\quad\left.\times\hat{\Lambda}_j(q)\left[1+\frac{\hat{n}[\hat{\mathcal{A}}(\phi_+)-\hat{\mathcal{A}}'(\phi_+)\phi_-/2]}{2p'_-}\right](\hat{p}'+m)(1+s'\gamma^5\hat{s}')\right\},
\end{split}
\end{equation}
where
\begin{equation}
\bm{\pi}_{\perp,e}(\phi)=\frac{\bm{p}_{\perp}}{m}-\frac{p_-}{q_-}\frac{\bm{q}_{\perp}}{m}-\frac{\bm{\mathcal{A}}_{\perp}(\phi)}{m}
\end{equation}
and $\bm{\mathcal{E}}(\phi)=-\bm{\mathcal{A}}'(\phi)$. Due to the inverse scaling of the damping terms with the minus components of the momenta in Eq. (\ref{P_NCS}), one may think that a first-order expansion in $\phi_-$ is required for those terms. However, one realizes that the resulting terms would provide a local correction in $\phi_+$ of the order of $\alpha$ to the term linear in $\phi_-$ in the phase, which can be neglected within our leading-order treatment.

Equation (\ref{P_NCS_f}) can be further manipulated and the integrals in $d\phi_-$ and in $d^2\bm{q}_{\perp}$ can be taken analytically with standard methods, as it has been carried out, e.g., in Refs. \cite{Di_Piazza_2018,Di_Piazza_2019}: the integral in $d\phi_-$ results in modified Bessel functions and the integral in $d^2\bm{q}_{\perp}$ is Gaussian. Also the Dirac trace is easily computed and spin- and polarization-resolved probabilities of Compton scattering are known \cite{Baier_b_1998,Seipt_2018,Li_2019,Chen_2019,Li_2020,Seipt_2020}. Our main remark here is that these results can still be used here. The novelty of the present analysis is, however, that the decay of the electron and the photon states affects the total emission probability and not simply as an overall damping factor (this was already found within the probabilistic approach in Ref. \cite{Tamburini_2019}). The precise structure of the decaying exponential functions depends on whether the corresponding particle is either an incoming or an outgoing particle. In the case of an incoming (outgoing) particle in a given state and for a fixed light-cone time $\phi_0$, the decay exponent corresponds to the total probability that the particle in that state radiates a photon via nonlinear Compton scattering (for electrons and positrons) or transforms into an electron-positron pair via nonlinear Breit-Wheeler pair production (for photons) from $\phi\to-\infty$ to $\phi=\phi_0$ (from $\phi=\phi_0$ to $\phi\to\infty$). For this reason, we can conclude that for $\chi_0\sim 1$, the effects of the particles states decay are important for pulse phase lengths $\Phi_L$ such that $\alpha\xi_0\Phi_L\gtrsim 1$ \cite{Ritus_1985,Baier_b_1998}. 

It is also important to stress that the decaying exponential depend on all quantum numbers characterizing the particles, i.e., not only on their momentum but also on their spin (for electrons and positrons) and polarization (for photons). These spin- and polarization-effect significantly complicate the computations as they prevent using the well-known spin and polarization sum-rules that simplify the evaluation of the Dirac traces in perturbative QED \cite{Landau_b_4_1982,Peskin_b_1995}. 

Finally, we observe that if we ignore electron spin-effects and photon polarization-effects (i.e., use probabilities averaged (summed) over the discrete quantum numbers of the initial (final) particles) as well as the decay of the photon state, our results are in agreement with those in Ref. \cite{Tamburini_2019}.

We pass now to the case of nonlinear Breit-Wheeler pair production. In this case, the leading-order amplitude in $\alpha$ but taking into account the decay of the states is given by
\begin{equation}
S^{(\gamma\to e^-e^+)}=-ie\int d^4x\,\frac{\bar{U}^{(\text{out})}_{R,s'}(p',x)}{\sqrt{2\varepsilon'}}\frac{\hat{\mathscr{A}}^{(\text{in})}_{R,j}(q,x)}{\sqrt{2\omega}}\frac{V^{(\text{out})}_{R,s}(p,x)}{\sqrt{2\varepsilon}},
\end{equation}
where, for notational convenience, we have assumed that the positron is produced with four-momentum $p^{\mu}$ and spin quantum number $s$. In fact, due to the symmetry structure of the amplitude, we can already conclude that the probability of nonlinear Breit-Wheeler pair production within the LCFA and by taking into account the decay of the states is given by (see also Ref. \cite{Di_Piazza_2019} for the expression of the probability without the decay of the states)
\begin{equation}
\label{P_NBWPP_f}
\begin{split}
&P^{(\gamma\to e^-e^+)}=\int\frac{d^3p}{16\pi^2}\frac{\alpha}{q_- p'_-\varepsilon}\int d\phi_+e^{2\text{Im}\left\{\frac{m}{q_-}\int_{-\infty}^{\phi_+}d\varphi P_j(q,\varphi)+\int_{\phi_+}^{\infty}d\varphi \left[\frac{m}{p'_-}M_{s'}(p',\varphi)+\frac{m}{p_-}M_s(-p,\varphi)\right]\right\}}\\
&\quad\times\int d\phi_- e^{i\frac{m^2}{2p_-}\frac{q_-}{p'_-}\left\{[1+\bm{\pi}_{\perp,p}^2(\phi_+)]\phi_-+\frac{\bm{\mathcal{E}}^2(\phi_+)}{m^2}\frac{\phi_-^3}{12}\right\}}\frac{1}{4}\text{tr}\left\{\left[1-\frac{\hat{n}[\hat{\mathcal{A}}(\phi_+)+\hat{\mathcal{A}}'(\phi_+)\phi_-/2]}{2p'_-}\right]\hat{\Lambda}_j(q)\right.\\
&\quad\times\left[1-\frac{\hat{n}[\hat{\mathcal{\mathcal{A}}}(\phi_+)+\hat{\mathcal{A}}'(\phi_+)\phi_-/2]}{2p_-}\right](\hat{p}-m)(1+s\gamma^5\hat{s})\left[1+\frac{\hat{n}[\hat{\mathcal{A}}(\phi_+)-\hat{\mathcal{A}}'(\phi_+)\phi_-/2]}{2p_-}\right]\\
&\quad\left.\times\hat{\Lambda}_j(q)\left[1+\frac{\hat{n}[\hat{\mathcal{A}}(\phi_+)-\hat{\mathcal{A}}'(\phi_+)\phi_-/2]}{2p'_-}\right](\hat{p}'+m)(1+s'\gamma^5\hat{s}')\right\},
\end{split}
\end{equation}
where
\begin{equation}
\bm{\pi}_{\perp,p}(\phi)=\frac{\bm{p}_{\perp}}{m}-\frac{p_-}{q_-}\frac{\bm{q}_{\perp}}{m}+\frac{\bm{\mathcal{A}}_{\perp}(\phi)}{m}
\end{equation}
and where we have also used the negative-energy electron density matrix $v_s(p)\bar{v}_s(p)=(\hat{p}-m)(1+s\gamma^5\hat{s})/2$ \cite{Landau_b_4_1982}. Analogous remarks about the importance of the particles states decay effects and of the dependence of the decay exponential functions on the electron/positron and photon quantum numbers apply here too.

\section{Conclusions}
In conclusion, we have derived analytical expressions of the probability of nonlinear Compton scattering and nonlinear Breit-Wheeler pair production within the locally-constant field approximation by including the effects of the decay of the particles states but still neglecting radiative corrections of the order of $\alpha$. The effects of the decay of the states, in fact, are cumulative effects scaling with the laser pulse duration and amount to exponential damping factors, which take into account the fact that electron/positron and photon states are not stable states in a plane wave. 

After solving the Schwinger-Dyson equations for electron, positron, and photon in- and out-states, we have inserted them into the leading-order in $\alpha$ amplitudes of nonlinear Compton scattering and nonlinear Breit-Wheeler pair production to determine the effects of the states decay into the corresponding probabilities. We have found that these probabilities can be expressed as integrals over the light-cone time $\phi$ of the corresponding probabilities without states decay per unit $\phi$ times a light-cone time-dependent exponential damping function for each participating particle. The exponential function for an incoming electron/positron (photon) at each light-cone time corresponds to the total probability that the electron/positron (photon) emits a photon via nonlinear Compton scattering (transforms into an electron-positron pair via nonlinear Breit-Wheeler pair production) until that light-cone time. Analogously, the exponential damping function for an outgoing electron/positron (photon) at each light-cone time corresponds to the total probability that the electron/positron (photon) emits a photon via nonlinear Compton scattering (transforms into an electron-positron pair via nonlinear Breit-Wheeler pair production) from that light-cone time on. 

Interestingly, the exponential damping functions depend non only on the particles momentum but also on their spin (for electrons/positrons) and polarization (for photons). An important consequence of this last dependence is that the spin and polarization sum-rules employed in perturbative calculations cannot be used anymore. Finally, since the exponential damping functions feature light-cone-time-integrated probabilities of nonlinear Compton scattering and nonlinear Breit-Wheeler pair production (computed for stable particles), the effects of the states decay at $\chi_0\sim 1$ are expected to become significant when the laser pulse length is sufficiently large that $\alpha\xi_0\Phi_L\gtrsim 1$.

\acknowledgments{This publication is also supported and T. P. is funded by the Collaborative Research Centre 1225 funded by Deutsche Forschungsgemeinschaft (DFG, German Research Foundation) - Project-ID 273811115 - SFB 1225.}

%


\end{document}